\documentstyle[prb,psfig,epsfig,aps]{revtex}

\begin{document}
\draft

\input{epsf}

\title{
  AC induced damping of a fluxon in long Josephson junction
}
\author{ M.~V.~Fistul$^1$, E. Goldobin$^2$, and A.~V.~Ustinov$^3$}
\address{
  $^1$Max-Planck Institut f\"ur Physik Komplexer Systeme,
  D-01187 Dresden Germany
}
\address{
  $^2$Institute of Thin Film and Ion Technology (ISI),
  Forschungszentrum J\"ulich GmbH (FZJ),
  D-52425 J\"ulich, Germany
}
\address{
  $^3$Physikalisches Institut III,
  Universit\"at Erlangen-N\"urnberg,
  D-91058 Erlangen, Germany
}

\date{\today}

\wideabs{ %REVTeX 3.1 feature

\maketitle

\begin{abstract}

  We present a theoretical and experimental study of Josephson vortex
  (fluxon) moving in the presence of spatially {\it homogeneous} dc and
  ac bias currents. By mapping this problem to the problem of calculating
  the
  current-voltage characteristic of a small Josephson junction, we derive the
  dependence of the average fluxon velocity $u$ on the dc bias current
  $\gamma$. In particular we find that the low frequency ac bias current
  results in an additional nonlinear damping of fluxon motion. Such
  ac induced damping crucially depends on the intrinsic
  damping parameter $\alpha$ and increases drastically as $\alpha$ is
  reduced.  We find a good agreement of the analysis with both the direct
  numerical simulations and the experimentally measured $\gamma(u)$
  curves of a long annular Josephson junction with one trapped fluxon.

\end{abstract}


}%end of WideAbs

\section{Introduction}

Solitary waves have attracted a great attention for many years because they
can be excited and observed in diverse nonlinear systems.
\cite{Strogatz,Scott} Well-known examples of such solitary waves are optical
solitons, domain walls, kinks and magnetic fluxons, just to name a few.

The latter case, a fluxon carrying the magnetic flux quantum $\Phi_0$, has a
form of a {\it Josephson vortex} moving along a long
Josephson junction (LJJ). This system presents an additional interest
because the fluxon motion can be easily controlled by externally applied
magnetic field and dc bias.\cite{GEns,UMT,WFU} It is  also well known that
under these conditions the fluxon motion can be mapped to the classical
mechanics of a macroscopic (relativistic) particle and such correspondence
is practically exact.\cite{Scott,GEns,UMT,WFU}

It was shown theoretically and experimentally that in the presence of both
dc and ac bias the fluxon motion becomes more complex and it displays
various fascinating phenomena.
\cite{GEns,MSh,Malomed,Mal2,UMal,MFU} Thus, as the frequency
$\omega$ of ac bias is large, i.e. $\omega~\simeq~\omega_p$, where
$\omega_p$ is the plasma frequency of a LJJ, the fluxon radiates plasma
waves.\cite{Scott,MSh,Malomed} When the frequency $\omega $ is small
($\omega\ll\omega_p$) but {\it inhomogeneous} (static or dynamic) potential
is present, the fluxon motion can be locked to the external ac bias. In this
case the vertical steps (``Shapiro steps") appear in the dependence of the
average fluxon velocity $u$ on the dc bias $\gamma$ (analog of the $I$--
$V$ curve).\cite{GEns,Mal2,UMal} Moreover, in the same frequency region
the resonant escape of a pinned fluxon has been observed.\cite{MFU}

However, the fluxon motion in the presence of both {\it homogeneously}
applied dc and ac bias for the frequency $\omega\ll\omega_p$, has not
been investigated neither theoretically nor experimentally. In this Report we
present a theoretical and experimental study of this limit. We show that,
unexpectedly, the influence of a small ac bias is rather large, and an
{\it additional ac induced damping} increases as the standard damping
characterized by parameter $\alpha$ reduces.

\section{Equation of motion and theoretical analysis}

The dynamics of a fluxon in a long Josephson junction is described by a
perturbed sin-Gordon equation for the Josephson phase $\varphi (x,t)$.
\cite{Scott,Bar} For a particular case of homogeneous  dc ($\gamma$) and
ac ($\varepsilon \sin \omega t$) bias currents this equation has the form
\cite{MSh,Malomed}
\begin{equation}
  \varphi_{xx}-\varphi_{tt}-\sin\varphi
  = \alpha\varphi_{t} - \gamma - \varepsilon \sin\omega t.
  \label{GenEq}
\end{equation}
Here, the units of time and coordinate are correspondingly, the inverse
plasma frequency $1/\omega_p$ and the Josephson penetration length
$\lambda_J$. The dimensionless damping parameter
$\alpha$ is determined by the direct losses that are due to
the presence of quasiparticle (dissipative) current.

In the absence of perturbation (rhs of Eq.~(\ref{GenEq}) ) the solution of
Eq.~(\ref{GenEq}) is a a fluxon moving with velocity $u$.
In a general case we use the standard
transformation of the Eq.~(\ref{GenEq}) to the fluxon reference frame,
where fluxon moves with an {\it average} velocity $u$. New variables are
defined as
\[
  \xi  = \frac{x-ut}{\beta}~~,
  \tau = \frac{t-ux}{\beta}~~,
  \beta= \sqrt{1-u^2}.
\]
The Eq.~(\ref{GenEq}) rewritten in the new variables is
\begin{equation}
  \varphi_{\xi\xi}-\varphi_{\tau\tau}-\sin\varphi
  = -\frac{\alpha  }{\beta}\varphi_{\tau}
  +  \frac{\alpha u}{\beta}\varphi_{\xi}
  - \gamma - \varepsilon \sin\left( \omega \frac{\tau+\xi u}{\beta} \right)
  . \label{GenEq:NewCoor}
\end{equation}
Note here, that the fluxon, in its reference frame, ``feels'' the periodic
coordinate and time dependent potential given by the last term of
Eq.~(\ref{GenEq:NewCoor}).

We are interested in the regime of small frequencies of ac bias current,
namely $\omega \ll 1$ in the dimensionless units.  In this limit we can
neglect the interaction of the fluxon with the plasma  oscillations.
\cite{MSh,Malomed} The potential energy $E_{\rm pot}$ of  the fluxon is:
\begin{equation}
  E_{\rm pot}=\int\limits_{-\infty}^{+\infty}
  \left[
    \frac{1}{2}\varphi_\xi^2
    - \gamma\varphi + (1-\cos\varphi)
    - \varepsilon \varphi \sin\left( \omega\frac{\tau+\xi u}{\beta} \right)
  \right] d\xi
  .\label{PotEnergy}
\end{equation}
Introducing the fluxon solution in the form
$\varphi(\xi,\tau)=\varphi_0(\xi-X(\tau))$, where $\varphi_0(\xi)$ is a well
known Josephson vortex solution we derive the equation of motion for the
coordinate $X(\tau)$ of  the fluxon (in the reference frame):
\begin{equation}
  \ddot X+\frac{\alpha}{\beta} \dot X+\frac{\alpha u}{\beta}
  = \frac{\pi}{4}\gamma
  - \frac{\pi \varepsilon}
         {4 \cosh\left( \displaystyle\frac{u\omega \pi}{2\beta} \right)}
    \sin\left( \omega\frac{\tau+X u}{\beta} \right)
  .\label{EqMot}
\end{equation}
Thus, the main result can be seen from this equation: the applied ac bias
current (last term in Eq. (\ref{EqMot}) )
excites fluxon librations of the frequency $\frac{\omega}{\beta}$
and, in turn, the interaction of such librations
with the ac bias current leads to the additional nonlinear damping of a
fluxon motion.
The interesting feature of the Eq.~(\ref{EqMot}) is that it can be mapped to
the problem of calculating the $I$--$V$ characteristic (IVC) of the small
Josephson junction with a specific damping, in general, different from
$\alpha$. To see a precise mapping we introduce new variables:
\[
  y = \frac{u\omega}{\beta} X , \quad
  \tau_0 = \tau \frac{\omega}{\beta},
\]
and obtain the equation
\begin{equation}
 \ddot y(\tau_0) + \frac{\alpha}{\omega}\dot y(\tau_0)
 + i_c(u)\sin(y+\tau_0 )
 = \frac{u\beta}{\omega}
 \left( \frac{\pi\gamma}{4} - \frac{\alpha u}{\beta} \right)
 ,\label{EqMot2}
\end{equation}
where $i_c(u)$ is a ``critical current'' of the small junction which depends
on the velocity $u$ of the fluxon from our problem and is given by:
\begin{equation}\label{EqMot2:ic}
  i_c(u) = \frac{\pi u \beta \varepsilon}
  {4\omega\cosh\left( \displaystyle\frac{u\omega \pi}{2\beta} \right)}
\end{equation}
The $\gamma$--$u$ characteristics of moving fluxon can be found from this
equation in the following form:
\begin{equation}
  \gamma = \frac{4\alpha u}{\pi\beta}
  + \frac{4 \omega}{\pi u \beta} \left[ i(1)-\frac{\alpha }{\omega} \right]
  ,\label{I-Vcurve}
\end{equation}
where $i(v)$ is the nonlinear
{\it current-voltage characteristics of a small Josephson
junction} with the critical current $i_c(u)$ and the damping
$\alpha/\omega$. In
Eq.~(\ref{I-Vcurve}), the first term is the usual $\gamma$--$u$
characteristic for a fluxon moving along LJJ and the second term
corresponds to the correction due to the presence of ac bias current.

\begin{figure}
\centering  \psfig{figure=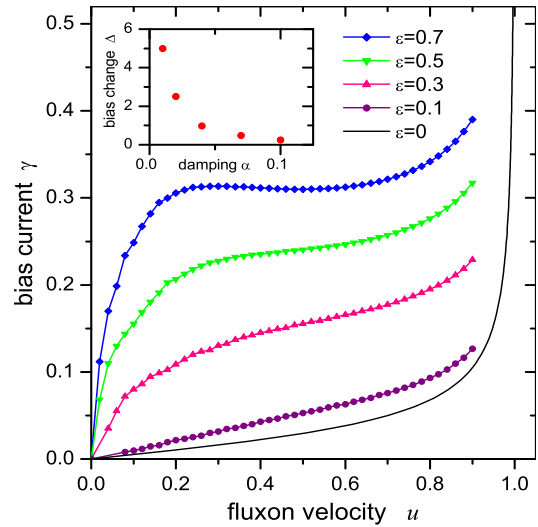 ,height=7cm ,width=7cm}
  \vspace{2mm}
%\centerline{\epsfig{file=fig1.eps,width=3in}}
  \caption{
    The $\gamma$--$u$ curves obtained by using Eq. (\ref{I-Vcurve})
    for various values of the ac bias amplitudes:
    $\varepsilon~=~0, ~0.1, 0.3, ~0.5, ~0.7$.
    The damping parameter $\alpha$ and ac frequency
    $\omega$ are, correspondingly, $0.04$ and $0.03$. The inset shows the
    normalized dc bias change $\Delta$ to the applied ac bias
    $\varepsilon=0.1$ at $u_0=0.3$, as a function of the damping  parameter
    $\alpha$.
  }
  \label{Fig:gamma(u)}
\end{figure}

The analytical expression for $i(v)$ dependence for small JJ at arbitrary
values of damping and critical current is not known. Using numerically
simulated $i(v)$
\cite{comment1} and Eq.~(\ref{I-Vcurve}), we obtain the $\gamma(u)$
dependence for various values of $\varepsilon$. Fig.~\ref{Fig:gamma(u)}
shows $\gamma$--$u$ characteristics calculated in this way for
$\varepsilon$ between zero and $0.7$.

Let us analyze some limits. For small fluxon velocity, namely
$u\ll\max{\{\alpha, \omega \}}/\varepsilon$, $\gamma(u)$ looks like a straight
line, but its slope is steeper than in the autonomous case $\varepsilon=0$
(see Fig.~\ref{Fig:gamma(u)}):
\begin{equation}
  \gamma = \frac{4\alpha u}{\pi \beta}
  \left[
    1+\frac{1}{2}\left(\frac{\pi\varepsilon}{4}\right)^2
    \frac{\beta^2}{\alpha^2+\omega^2}
  \right]
  .\label{I-Vcurve:Small}
\end{equation}
In the opposite case $u\gg\max(\alpha, \omega)/\varepsilon$, the
$\gamma$-- $u$ dependence just shifts by the constant value
$\varepsilon\leq1$ from the autonomous curve. Thus, ac bias induced
damping is
highly nonlinear, and
the $\gamma$-- $u$ dependence displays an inflection
point for moderate values of $u$ (see Fig. 1).

In the region of small frequencies $\omega \simeq \alpha$ as the damping
coefficient $\alpha$ decreases, the response of the fluxon to the ac bias
current (which is proportional to $1/\alpha^2$) drastically increases
[see Eq.~(\ref{I-Vcurve:Small})].
In order to analyze this effect in more precise terms
we calculate the change of dc bias
$\Delta=\frac{\gamma(u_0)-\gamma_0(u_0)}{\gamma_0(u_0)}$
as a function of the damping parameter $\alpha$,
for particular values of fluxon velocity $u_0~=~0.3$ and ac bias
$\varepsilon~=~0.1$, where $\gamma_0(u_0)$
is the value of dc bias current in the absence of ac bias.
The results are presented in the inset of Fig.~\ref{Fig:gamma(u)}.

\section{Numerical simulations}

In order to verify the analysis presented above and predictions made we
performed direct numerical integration of the PDE (\ref{GenEq}). The
periodic
boundary conditions
\begin{eqnarray}
  \varphi(\ell,t) = \varphi(0,t)+2\pi
  ; \label{Eq:BC_1}\\
  \varphi_x(\ell,t) = \varphi_x(0,t)
  , \label{Eq:BC_2}
\end{eqnarray}
were imposed and the normalized length of the LJJ $\ell$ was taken to be
$10$. We used the damping coefficient $\alpha=0.04$ and the ac bias
frequency $\omega = 0.03$ as these values are very close to the
experimental ones (see the next section). The obtained $\varphi(x,t)$
profile was used to find the voltage (proportional to the fluxon velocity
$u$) across the junction. The averaging of velocity was performed over the
time interval which is a multiple of the period of the ac frequency. We used
two programs with different numerical schemes and averaging algorithms,
and obtained identical results.

\begin{figure}
%\centering  \psfig{figure=fig2.eps ,height=7cm ,width=3in}
  \centerline{\epsfig{file=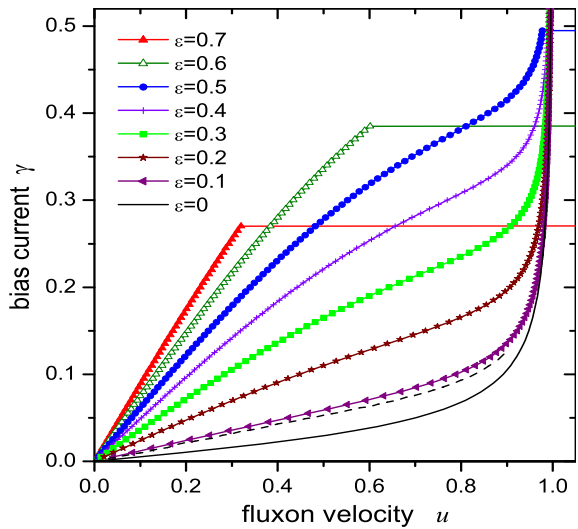,height=7cm,width=3in}}
  \vspace{2mm}
  \caption{Numerically simulated dc driving force vs velocity $\gamma(u)$
  curves for various amplitudes of ac drive $\varepsilon$ indicated in the
legend. For comparison, dashed line shows the theoretical curve for
$\varepsilon~=~0.1$ taken from Fig.~\protect\ref{Fig:gamma(u)}.}
 \label{Fig:numer-gamma(u)}
\end{figure}

The $\gamma$--$u$ curves for different values of the ac bias amplitude
$\varepsilon$ are shown in Fig.~\ref{Fig:numer-gamma(u)}.
The main effect of the ac induced damping which depends on the fluxon
velocity is clearly present in numerical data.

Moreover, as the ac bias is
small ($\varepsilon~\leq~0.2$) the qualitative
agreement between analytical results obtained above using the
perturbation theory and direct numerical simulation is rather good practically
in a whole region of fluxon velocities $u<1$
(see Fig.~\ref{Fig:numer-gamma(u)}).
However, as
$\varepsilon$ increases the theoretical analysis {\it overestimate}
the ac induced damping in the region of small $u$ and {\it underestimate} it
in the opposite regime of large $u$. Most probably, the reason for that is
that we neglect the interaction of the fluxon with the plasma oscillations.
The direct numerical simulations also show that
for high ac bias ($\varepsilon=0.5, 0.6,$ and $0.7$)
the state of moving fluxon becomes unstable,
and the simulated curves in their upper range end with a point of switching
to high voltages.

\section{Experiment}

To observe the predicted effect we carried out real measurements using
Nb/Al-AlO$_x$/Nb long annular Josephson junction \cite{Hypres} with a
trapped magnetic fluxon. The junctions were underdamped with the typical
value of $\alpha \approx 0.04$ at $T=4.2\,{\rm K}$. The junction has the
mean radius $R=46\,{\rm \mu m}$ and width $w=5\,{\rm \mu m}$. The
critical current density $j_c$ was about $100\,{\rm A/cm^2}$, which
corresponds to the normalized junction length $\ell$ of about 10. A single
magnetic fluxon (Josephson vortex) was trapped in the junction by applying
a small dc bias current during the transition from the normal to the
superconducting state while cooling down the sample.\cite{WFU,MFU}

\begin{figure}
  \centerline{\epsfig{file=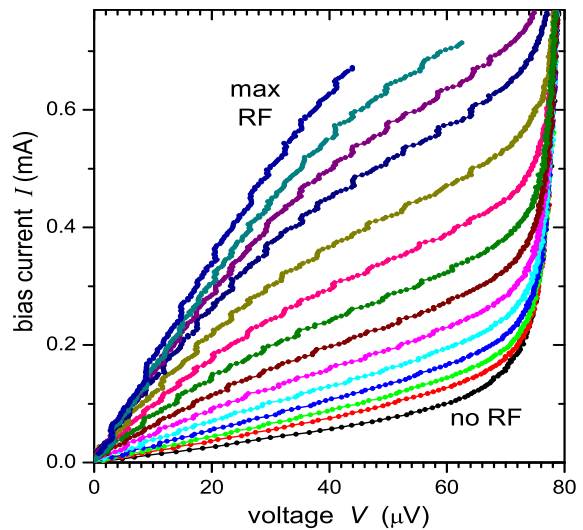,height=7cm,width=3in}}
 \vspace{2mm}
  \caption{The experimentally measured $I$-$V$ curves of an annular LJJ
  with trapped fluxon for different values of microwave ac drive of the
  frequency 1.4 GHz at power levels varying from 4.3 dBm to 20.6 dBm. The
  lowest curve corresponds to the autonomous $I$--$V$ curve (zero ac
  power). }
   \label{Fig:exp}
\end{figure}

For the external ac bias the microwave radiation with different frequencies
in the range from 1 GHz to 20 GHz was used. Here we present the results
only for $f\equiv\omega/(2\pi)=1.4\,{\rm GHz}$ which corresponds to the
$\omega \simeq 0.03 \omega_p$. The microwave signal was applied by an
antenna placed close to the sample, so that ac bias current was induced by
the antenna in the bias leads of the junction. The typical power $W$ of the
applied microwave radiation was ranging from few pW to several tens of
mW, referenced to the top flange of the sample holder.

We measured $I$-$V$ curves of the annular Josephson junction with one
trapped magnetic fluxon at different ac power levels $W$. These $I$-$V$
are shown in Fig.~\ref{Fig:exp}. The measured $I$-$V$ curves qualitatively
look very similar to the $\gamma(u)$ dependencies obtained for the fluxon
in Sections II and III. Clearly, ac induced voltage-dependent damping is
observed (cf. Figs.~\ref{Fig:gamma(u)} and \ref{Fig:numer-gamma(u)}).
\begin{figure}
  \centerline{\epsfig{file=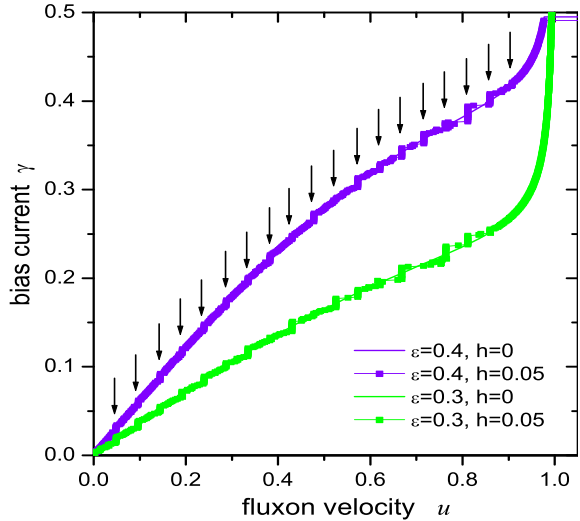,height=7cm,width=3in}}
 \vspace{2mm}
  \caption{
    The numerically simulated $\gamma$--$u$ curves in the presence of
    magnetic field $h=0.05$. The amplitude of ac drive was chosen
    $\varepsilon=0.3$ and $0.4$, and other parameters are the same as in
    Fig.~\protect\ref{Fig:gamma(u)}. Arrows identify Shapiro steps.
  }
     \label{Fig:comp_exp}
\end{figure}

\section{Discussion}

The $I$-- $V$ curves shown in Fig.~\ref{Fig:exp} display many small steps
which we explain by the presence of small non-uniformity of the LJJ. In fact
these steps are nothing else but conventional Shapiro steps induced by ac-
drive. The imperfection of the junction can be due to either self-field effects,
influence of the dc bias leads, or small technological inhomogeneities.

To check the influence of non-uniformity on simulated characteristics, we
performed additional numerical simulations with a tiny external magnetic
field $h$ applied in the junction plane. It is a well known that a such
magnetic field creates the coordinate dependent potential for the fluxon
\cite{GEns,UMT,WFU}, and therefore breaks the translational symmetry of
the system. This leads to the change of $\gamma$--$u$ curve such that the
fluxon rotation frequency in the annular system gets locked to the frequency
of ac drive in some ranges of $\gamma$. Physically, this effect corresponds
to the appearance of ac-induced voltage locking steps known as Shapiro
steps for Josephson junctions.

Figure~\ref{Fig:comp_exp} clearly shows that in the presence of magnetic
field there are numerous Shapiro steps of a small magnitude that appear in
the simulated $\gamma$--$u$ curve. Yet  the pronounced effect of ac
induced fluxon damping remains very evident.

\section{Conclusions}

We have reported a theoretical and experimental study of the fluxon motion
in the presence of dc and ac bias. By making use of the analysis and the
direct numerical simulations we obtain the ac induced nonlinear damping
of a fluxon. We also measured the current-voltage characteristics of an
annular LJJ with the trapped fluxon. We observed a such ac induced
damping as a large ascent of the $I$-$V$ curve in the presence of
externally applied microwave radiation. Because the response to the ac
bias increases as the intrinsic damping parameter $\alpha$ reduces
(Eq. (\ref{I-Vcurve:Small}) and Fig. 2), the observed effect can be useful as
a method to detect weak microwave radiation. In addition, this effects allows
to accurately measure the damping parameter $\alpha$ when it gets
extremely small at low temperatures.

\section{Acknowledgments}

We thank S.~Flach and B.~A.~Malomed for useful discussions,
A.~E.~Miroshichenko for help in the numerical analysis of
Eq. (\ref{EqMot2}) and A.~Wallraff for assistance in experiment.


\begin{references}
\bibitem{comment1} In order to obtain the $i(v)$ curve of a small
Josephson junction for different values of $i_c(u)$ we use two methods:
direct integration of the Eq. (\ref{EqMot2}) and Newton scheme allowing to
find the time periodic states. Both methods lead to the same results.
\end{references}
\end{document}